# The game of the "very small" and the "very big": The case of the Doppler Effect.

Bernhard Rothenstein, Physics Department, University of Timisoara, 1900, Timisoara, Romania, bernhard_rothenstein@yahoo.com
Albert Rothenstein, Centre for Vision Research, York University, Toronto, Canada

*Abstract.*
*The Doppler Effect is associated with physical quantities such as period, frequency, and wavelength. It involves a periodic phenomenon taking place at a given point in space and an agent which carries information about it in space. The device which creates the periodic phenomenon represents a source. The effect is observed from an inertial reference frame K(XOY) relative to which source S and receiver R move with speeds $v_s$ and $v_R$, respectively, whereas the information propagates with speed U. In order to keep the problem as simple as possible, we will consider the source as being at rest.*
*Moreau introduces the concept of non-locality in the period measurement by an observer, associated with the fact that a period measurement by a moving observer (accelerating or not) "requires the detection of (at least) two space-time events separated by a finite time interval, corresponding to the passage of successive wave-crests by the observer's detector" [1]. We can say that the moving observer R receives two successive wave-crests at two different points in space, at two different times. The time interval between the emissions of two successive wave-crests by the source, measured in its rest frame represents the period of emission, whereas the time interval between the receptions of the same two wave-crests by the receiver, measured in its rest frame represents the reception period. The problem is to find a relationship between the two periods defined above. It is clear that, in the general case, the speed of the receiver and the angle at which it receives the successive wave-crests could change. The important conclusion is that a continuous recording of the period of reception is not possible.*
*In many cases the non-locality is not taken into account. We present two cases in which a continuous change in the receiver's speed and in the angle at which the successive wave-crests are received takes place. In each case the error committed by not taking into account the non-locality is evaluated.*

1. Introduction

When speaking of a physical quantity it is advisable to specify who is performing the measurements, when and where the measurements are performed, the measuring devices used and the time taken to perform the measurements.
A Doppler effect experiment involves, in its simplest form, a stationary source of periodic oscillations S located at the origin O of the K(XOY) reference frame, the
rest frame of the medium through which information about the oscillations propagates with constant velocity U. It also involves a receiver R moving with constant velocity v along an arbitrary direction (Figure 1). We have to compare the constant period $T_S$ (the constant frequency $f_S=T_S^{-1}$) at which the source emits successive crests and the time interval $T_R$ (the frequency $f_R=T_R^{-1}$) during which they are received by R.

Moreau [1] introduces the concept of non-locality in the frequency (period) measurement by a moving observer. It is associated with the fact that a moving observer is not able to receive two successive crests at the same point in space.

Many derivations of the Doppler shift formulas, frequently used in the current practice, are based on simplifying assumptions (plane wave approximation or very small period approximation) which favor locality in the period measurement by a moving observer [2],[3],[4]. Such derivations go as follows. Consider that when clock $C_0(0,0)$ located at O reads $t_S$ a signal propagating with constant velocity U is emitted. The signal is received by R at a position $R(\theta,r)$ when a clock $C(\theta,r)$ located there reads $t_R$.

It is obvious that

$$t_R = t_S + \frac{r}{U} \tag{1}$$

Differentiating both sides of Equation (1) we obtain

$$dt_R = dt_S + \frac{dr}{U} \tag{2}$$

Equation (1) holds for any value of $t_S$, even for $t_S=0$.
Equation (2) leads to

$$\frac{dt_R}{dt_S} = \frac{1}{1 - \frac{v}{U}\cos\theta} \tag{3}$$

by taking into account that

$$\frac{dr}{dt_R} = v\cos\theta \tag{4}$$

represents the radial component of the instantaneous velocity of R and $\theta$ the instantaneous angle made between v and U (Figure 1)

If we think of $dt_S$ and $dt_R$ as of the "very small" periods at which the wave crests are emitted and received respectively, then Equation (3) describes the Doppler shift with stationary source and arbitrarily moving receiver with the locality assumption.

Many authors present it as a relationship between finite periods of emission and reception, without mentioning that it holds true only if the locality conditions are met [2],[3],[4].

Equation (4) shows that the measurement of the instantaneous velocity requires the same time interval $dt_R$ as the measurement of the reception time interval.

Following a derivation proposed by Donges [2] we consider that the phase of the oscillations at the source is

$$\psi_R = \omega_S t \tag{5}$$

whereas at the location of observer R it is

$$\psi_R = \omega_S\left(t - \frac{r}{U}\right) \tag{6}$$

$\omega_S$ representing the circular frequency of the oscillations taking place at the source.

Taking into account that, by definition, the time derivative of the phase equals the instantaneous frequency [5], we obtain

$$\omega_R = \omega_S\left(1 - \frac{v}{U}\cos\theta\right) \tag{7}$$

Equation (7) is frequently used even when it is considered that the instantaneous frequency is not a measurable physical quantity [5].

We could reduce the Doppler effect to its essence, replacing the crests by pellets fired by a machine gun at a constant period $T_S$, which hit a moving target at the time intervals $T_R$ [6]. It is clear that in this case the concept of instantaneous frequency and the derivations presented above are meaningless. Given this scenario, we have to find the time interval between the incidences of two successive bullets $T_R$ ($f_R$) and to relate it to $T_S$ ($f_S$). If $T_S$ is measured as a difference between the readings of the same clock $C_o(0,0)$ located in front of the source, $T_R$ is measured as a difference between the readings of two clocks located where the bullets hit the target. If the velocity of R is not constant or if between the receptions of two successive crests the angle θ changes, then the relationship between $T_S$ and $T_R$ becomes a function of time.

The purpose of our paper is to derive exact formulas for the acoustic Doppler shift taking into account the non-locality and presenting the error introduced by not taking it into account.

2. Longitudinal acoustic Doppler Effect with stationary source and uniformly accelerating receiver.

Consider that at t=0 (reading of clock $C_o(0,0)$ ), observer R is at rest and located at the origin O of the K(XOY) reference frame, where the source S(0,0) is located at rest. The observer starts to move at that very moment with constant acceleration g in the positive direction of the OX axis. At time $t_R$, R is located at a distance

$$x = \frac{1}{2} g t_R^2 \qquad (8)$$

With the locality assumption we consider that at time t the instantaneous velocity of R is

$$v = g t_R \qquad (9)$$

Replacing that value in Equation (7) we obtain

$$\omega_R = \omega_S \left(1 - \frac{g t_R}{U}\right) \qquad (10)$$

In the non-locality assumption, we consider that the stationary source emits successive wave-crests at a constant period $T_S$, labeled 0,1,2…n and emitted at t=0, $t_1=T_S$, $t_2=2T_S$…,$t_n=nT_S$. The motion of the crest labeled n is described by

$$x = U(t - nT_S) \qquad (11)$$

From Equations (8) and (11) we obtain that the n-th crest is received by R at time

$$t_{R,n} = \frac{U}{g}\left(1 - \sqrt{1 - 2\frac{g}{U}nT_S}\right) \qquad (12)$$

The time interval between the receptions of two successively emitted crests is given by

$$T_{R,n-1,n} = t_{R,n} - t_{R,n-1} = \frac{U}{g}(A_{n-1} - A_n) = \frac{2T_S}{A_{n-1} - A_n} \qquad (13)$$

where

$$A_n = \sqrt{1 - \frac{2g}{U}nT_S} \qquad (14)$$

When the speed of R equals the propagation speed of the information U, the observer loses contact with the source. This event takes place at time

$$t_{R,N} = \frac{U}{g} \qquad (15)$$

where N represents the order number of the last received crest which was emitted at time

$$NT_S = \frac{U}{2g} \qquad (16)$$

Expressed as a function of N, Equation (14) can be represented as

$$A_n = \sqrt{1 - \frac{n}{N}} \qquad (17)$$

We are now able to define a local Doppler factor

$$D = \frac{\omega_R}{\omega_S} = 1 - \frac{gt_R}{U} = \sqrt{1 - \frac{n}{N}} \qquad (18)$$

In Equation (18) we consider a continuous variation of $t_R$ and n.
We can also define a non-local Doppler factor

$$D_{n-1,n} = \frac{T_S}{T_{R,n-1,n}} = \frac{1}{2}(A_n + A_{n-1}) \qquad (19)$$

The error resulting from not taking into account the locality is given by

$$\varepsilon\% = \left(1 - \frac{D}{D_{n-1,n}}\right) \cdot 100 \tag{20}$$

and its variation with n for a given value of N is represented in Figure 2.
As we see, the error is an increasing function of n, due to the fact that the velocity of R increases with n, which favors the non-locality.

3. Stationary source and uniformly moving observer, receiving the wave-crests at an oblique angle of incidence.

The scenario is presented in Figure 3. The stationary source S is located at the origin O of the K(XOY) reference frame. Observer R moves with constant velocity v parallel to the OX axis at a distance d from it. In the non-locality approach we consider that during the reception of two successive wave-crests the angle θ at which the wave-crests are received changes. A change in the radial component of the velocity of R also takes place and we will take into account these essential features.
In the non-locality approach we consider that a wave-crest emitted at t=0 is received by R at a position $R_1(r_1,\theta_1)$ at time $t_{R,1}$. A second wave-crest emitted at $T_S$ is received by R at a position $R_2(r_2, \theta_2)$ at time $t_{R,2}$. We have

$$T_R = \frac{1}{U}(r_2 - r_1) + T_S \tag{21}$$

where $T_S$ represents the constant period at which the wave-crests are emitted by the source, whereas $T_R$ represents the time interval during which they are received by R at two different points in space. During the time interval $T_R$, R covers a distance $vT_R$. As we see

$$r_1^2 = r_2^2 + v^2 T_R^2 - 2r_2 v T_R \cos\theta_2 \tag{22}$$

Eliminating $r_1$ by combining Equations (21) and (22) and using the simplifying notations $r_1 = r$ and $\theta_2 = \theta$ we obtain

$$T_R^2\left(1 - \frac{v^2}{U^2}\right) - 2T_R\left[T_S + \frac{r}{U}\left(1 - \frac{v}{U}\cos\theta\right)\right] + T_S\left(T_S + \frac{2r}{U}\right) = 0 \tag{23}$$

Solved for $T_R$ equation (23) leads to a non-local Doppler factor

$$D_m = \frac{T_S}{T_R} = \frac{\omega_R}{\omega_S} = \frac{1 - \frac{v^2}{U^2}}{A - \sqrt{A^2 - B}} \tag{24}$$

where

$$A = 1 + \frac{d}{\lambda \sin\theta}\left(1 - \frac{v}{U}\cos\theta\right) \tag{25}$$

and

$$B = \left(1 - \frac{v^2}{U^2}\right)\left(1 + \frac{2d}{\lambda \sin\theta}\right) \tag{26}$$

Where $\lambda = UT_S$ represents the wavelength in the rest frame of the source. As we see, $D_m$ depends on the dimensionless parameter $m = d/\lambda$, d representing the distance between the line along which R moves and the OX axis. The locality assumption (very small period, very large source-receiver distance) is associated with very large values of m. As expected

$$\lim_{m \to \infty} D_m = D \tag{27}$$

where in accordance with Equation (7) D represents the Doppler factor corresponding to the locality assumption. The error made by not taking into account the non-locality can be define as

$$\varepsilon\% = \left(1 - \frac{D}{D_m}\right) \cdot 100 \tag{28}$$

The parameter $m = r/\lambda$ can change over a very large range. The ratio of v to U can also change over a large range, $v/U < 1$ corresponding to the subsonic case, $v/U = 1$ to the sonic case, whereas $v/U > 1$ corresponds to the supersonic case.
In order to illustrate the results obtained above we represent in Figure 4 the variation of the non-local Doppler factor D with the angle $\theta$ for a constant value of v/U but different values of the parameter m in the subsonic, sonic and supersonic cases. The case $v/U < 1$ correctly reflects the case when the source emits light signals,
The only thing that has to be done is to replace U with c and to impose the condition $v < c$. In the subsonic case, the variation of the error $\varepsilon\%$ introduced by not taking into account the non-locality with the angle $\theta$ is represented in Figure 5. As we can see, the error is a decreasing function of m, reaching a maximum in the range of the transversal Doppler Effect ($\theta = 90^0$).
For $0 < \theta < 90^0$ observer R is "outgoing" whereas for $90^0 < \theta < 180^0$ observer R is "incoming".

   4. Conclusions

We have shown that the derivation of the Doppler shift formulas is associated with the concept of locality (the moving observer is able to receive two successive crests while remaining located at the same point in space) or with the concept of non-locality (the moving observer receives two successive crests at two different points in space). The locality approach leads to the concept of simultaneous frequency defined as the time derivative of the phase. The formulas obtained in this case are so frequently used that authors fail to remind the reader that they are the result of the "plane wave approximation" or of the "very small period approximation." The non-locality approach is based on the time interval between the receptions of two successive crests, the inverse of which is defined as the reception frequency.

The use of the instantaneous frequency in cases in which the locality conditions are not met leads to significant errors, especially considering the high degree of accuracy with which time intervals are measured nowadays.

Figure captions

Figure 1. Scenario for deriving the Doppler shift formula in the case of the locality assumption. R represents the instantaneous position of an observer R in the rest frame of the source.

Figure 2. The error ε% introduced by not taking into account the non-locality.

Figure 3. Observer R moves with constant velocity parallel to the OX axis at a distance d from it. It receives two successive crests at two different points at different incidence angles.

Figure 4
Figure 4.a The variation of the non-local Doppler factor Dm with the incidence angle θ for different values of the parameter m=r/λ in the subsonic case v/U=0.5.
Figure 4.b The variation of the non-local Doppler factor Dm with the reception angle θ for different values of the parameter m=r/λ in the sonic case v/U=0.999.
Figure 4.c The variation of the non-local Doppler factor Dm with the reception angle θ for different values of the parameter m=r/λ in the supersonic case v/U=1.5.

Figure 5. The error introduced by not taking into account the non-locality as a function of the incidence angle θ for different values of the parameter m=r/λ in the subsonic case v/U=0.5.

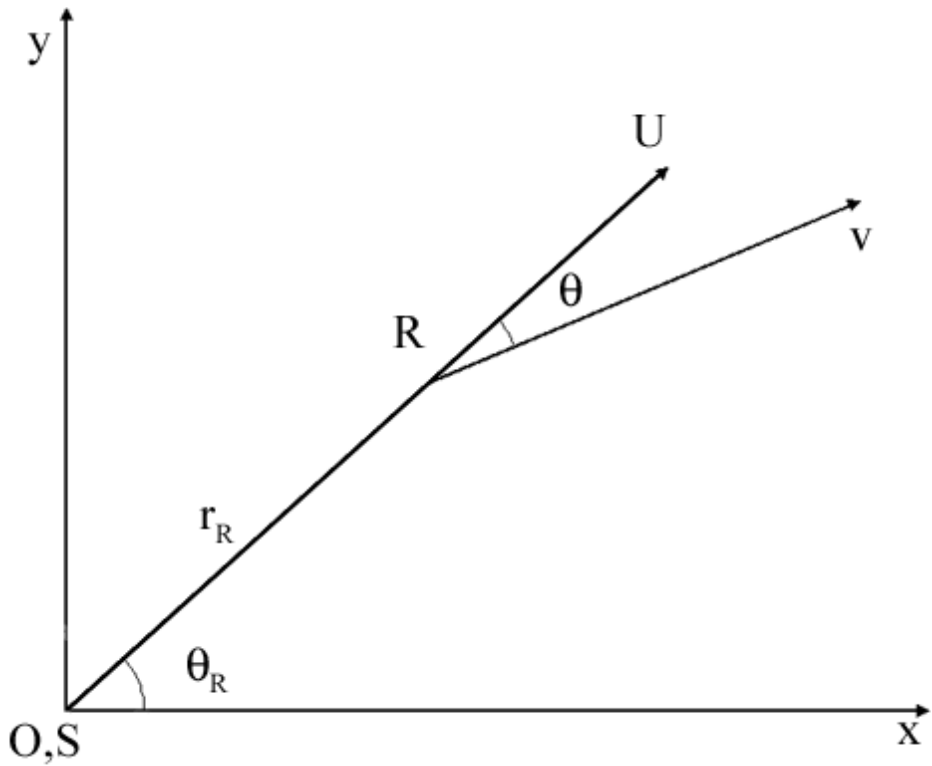

**Figure 1**

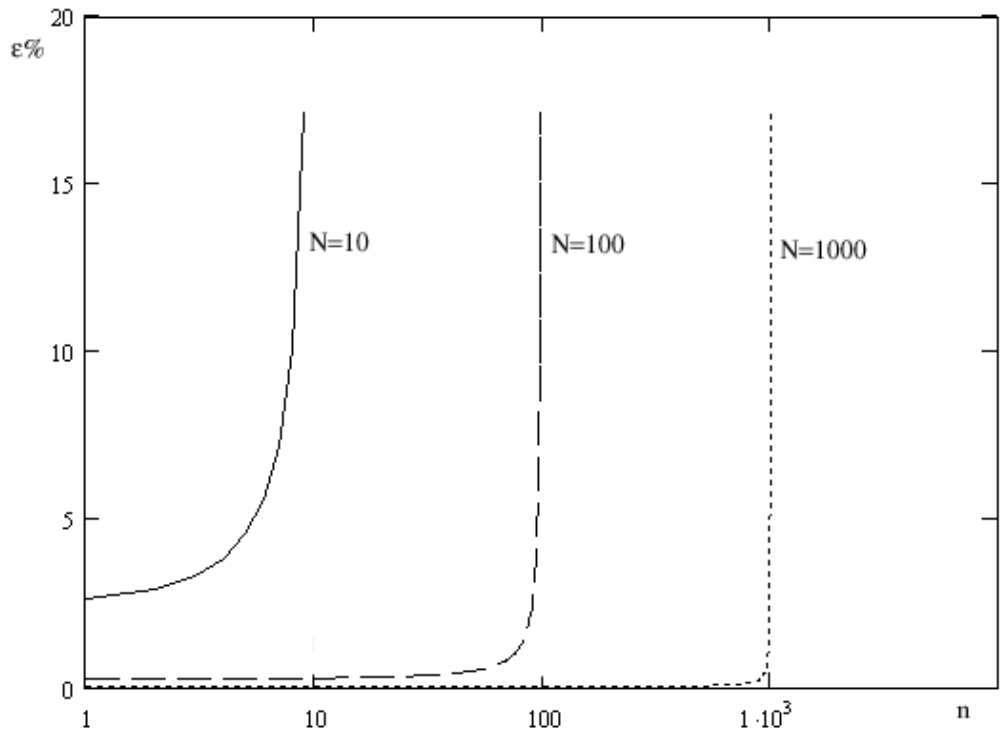

**Figure 2**

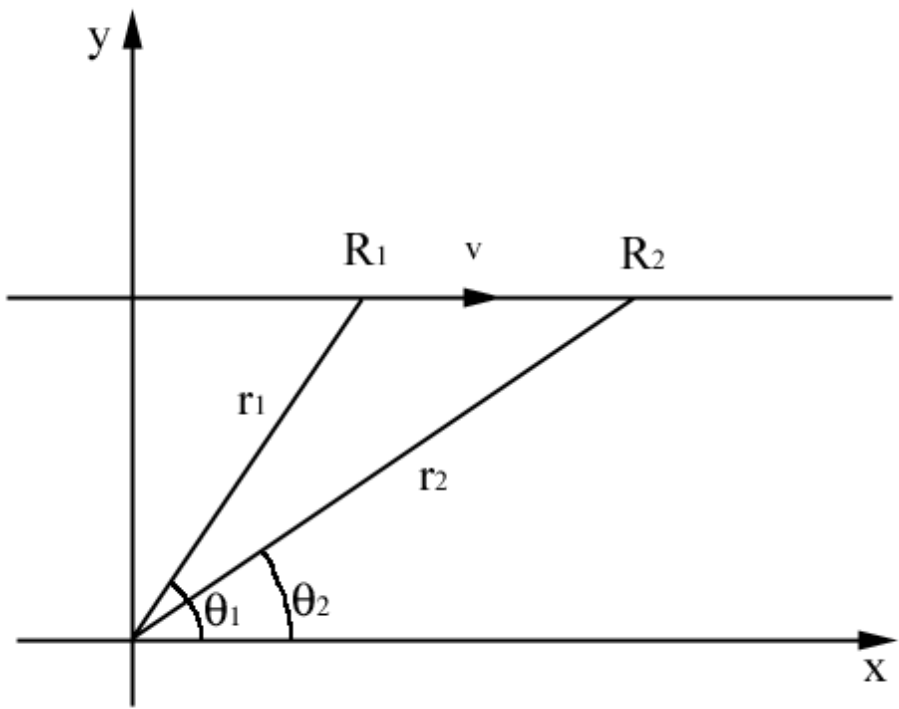

**Figure 3**

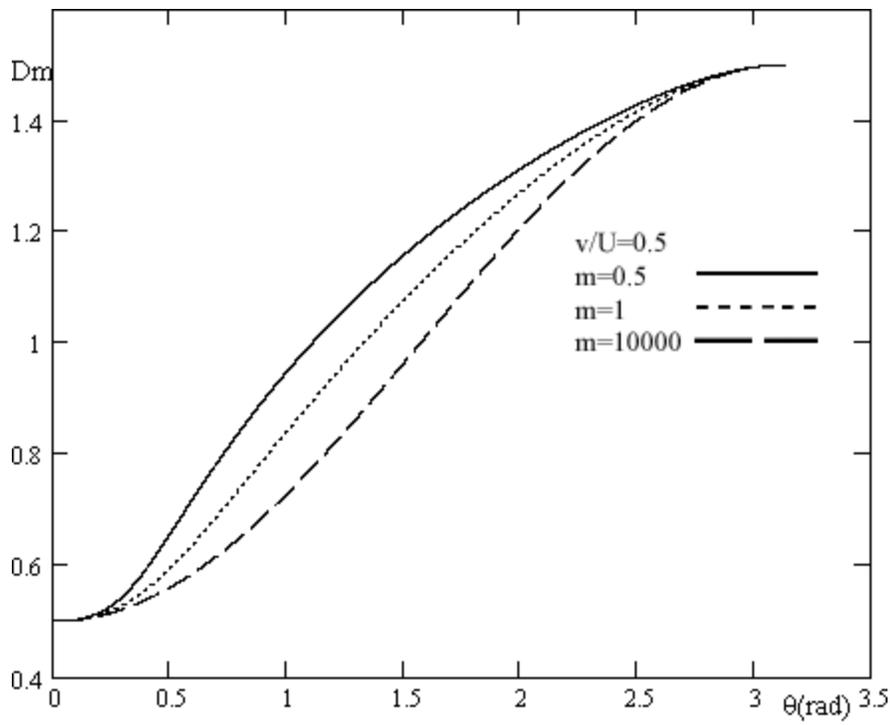

**Figure 4a**

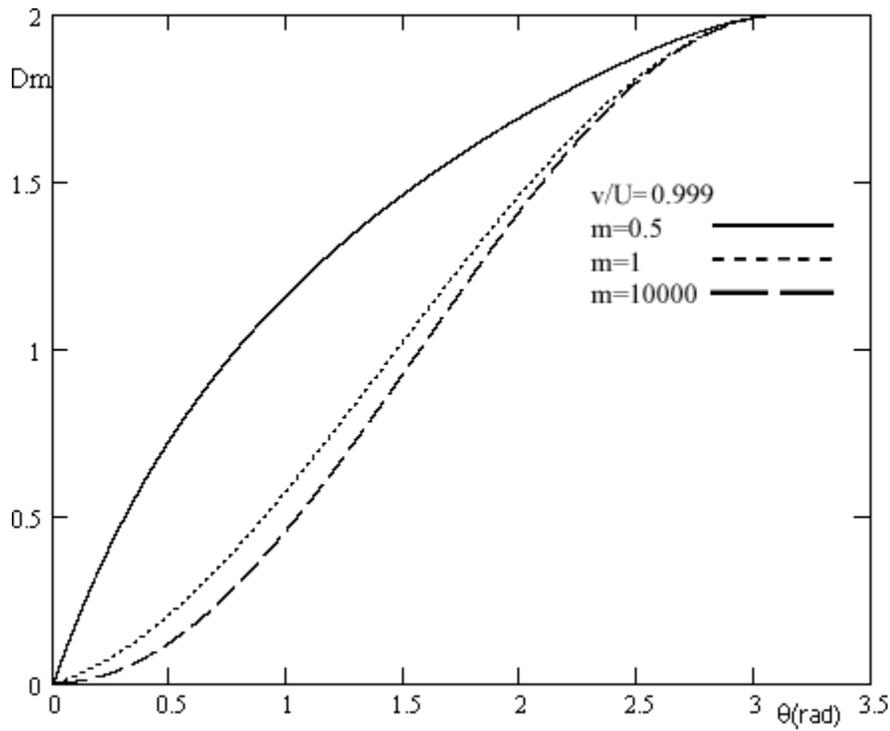

**Figure 4b**

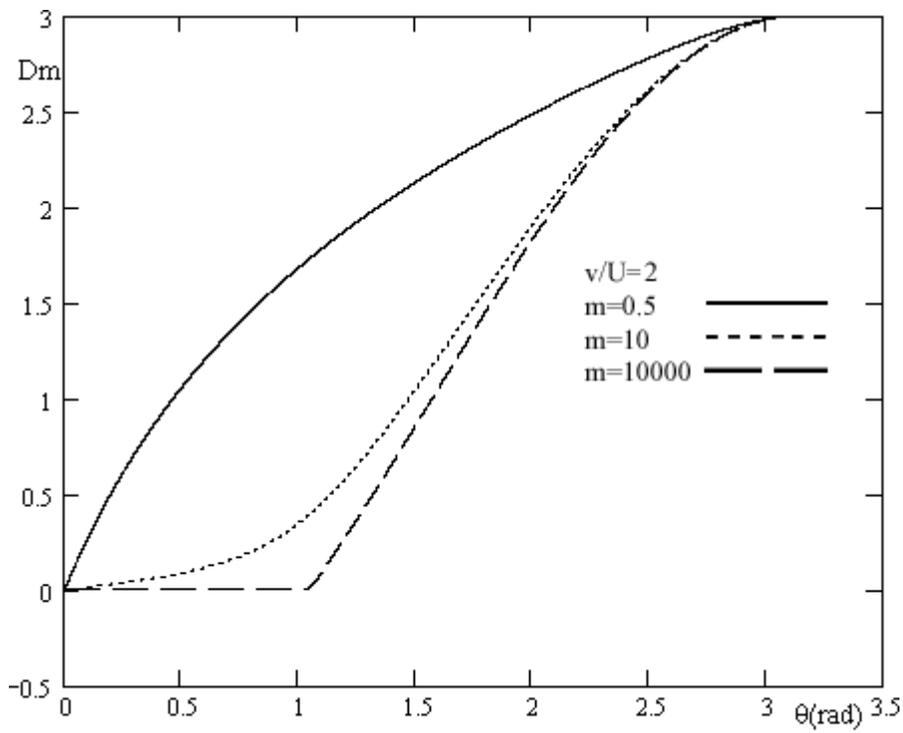

**Figure 4c**

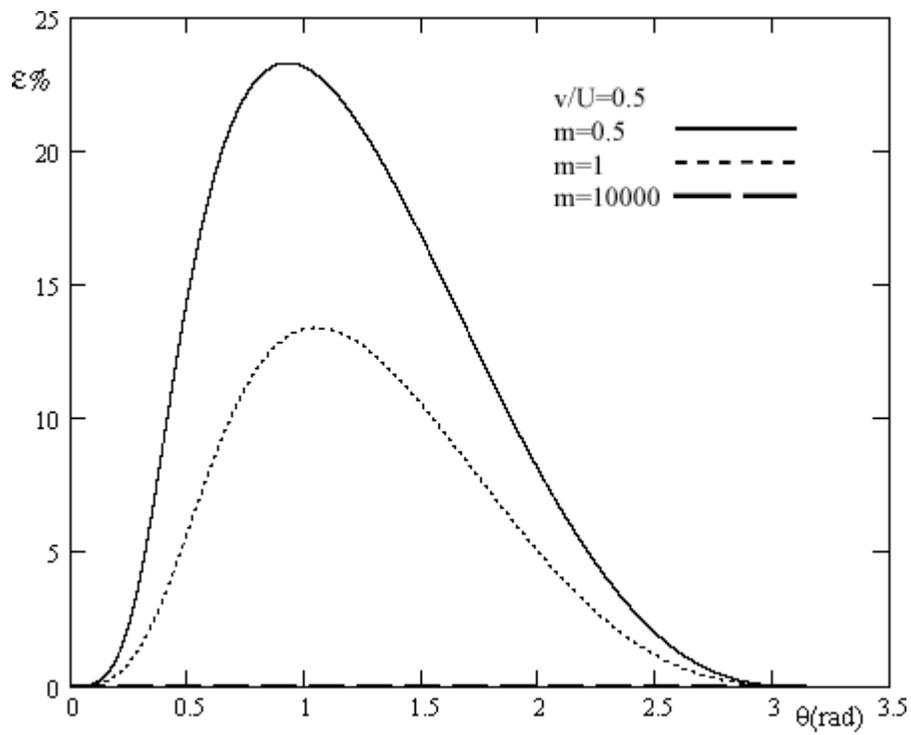

**Figure 5**